\documentclass[a4paper,twoside]{article}

\usepackage{epsfig}
\usepackage{subcaption}
\usepackage{calc}
\usepackage{amssymb}
\usepackage{amstext}
\usepackage{amsmath}
\usepackage{amsthm}
\usepackage{multicol}
\usepackage{pslatex}
\usepackage{apalike}
\usepackage{SCITEPRESS}     

\usepackage{hyperref}

\usepackage{comment}

\usepackage{listings}

\usepackage{mathtools}
\usepackage[all]{xy}
\usepackage{tikz}
\usepackage{multirow}
\usepackage{wasysym}
\usepackage{todonotes}

\usepackage{pgfplots}
\usepackage{pgfplotstable}
\pgfplotsset{compat=1.15}
\usetikzlibrary{pgfplots.groupplots}

\usepackage{import}
\usepackage{longtable}

\usepackage{adjustbox}
\usepackage{array}

\usepackage[binary-units=true]{siunitx}
\usepackage{graphicx}

\usepackage[utf8x]{inputenc}
\usepackage{tabularx}

\begin{document}

\title{Post Quantum Cryptography Analysis of TLS Tunneling On a Constrained Device}

\author{\authorname{Jon Barton, William J Buchanan\sup{1}\orcidAuthor{0000-0003-0809-3523}, Nikolaos Pitropakis\sup{1}\orcidAuthor{0000-0002-3392-9970}, Sarwar Sayeed\sup{1}\orcidAuthor{0000-0002-9164-7672}, Will Abramson}
\affiliation{\sup{1}Blockpass ID Lab, School of Computing, Edinburgh Napier University, Edinburgh, UK.}
\email{40327464@live.napier.ac.uk, \{b.buchanan, n.pitropakis, s.sayeed, will.abramson\}@napier.ac.uk}
}




\abstract{Advances in quantum computing make Shor's algorithm for factorising numbers ever more tractable.  This threatens the security of any cryptographic system which often relies on the difficulty of factorisation. It also threatens methods based on discrete logarithms, such as with the  Diffie-Hellman key exchange method. For a cryptographic system to remain secure against a quantum adversary, we need to build methods based on a hard mathematical problem, which are not susceptible to Shor's algorithm and create Post Quantum Cryptography (PQC). While high-powered computing devices may be able to run these new methods, we need to investigate how well these methods run on limited powered devices. This paper outlines an evaluation framework for PQC within constrained devices, and  contributes to the area by providing benchmarks of the front-running algorithms on a popular single-board low-power device. It also introduces  a  set  of  five  notions which can be considered to determine the robustness of particular algorithms. }

\onecolumn \maketitle \normalsize \setcounter{footnote}{0} \vfill

\section{\uppercase{Introduction}}\label{sec:introduction}
Public key (asymmetric) encryption methods are fundamental to the security of many digital systems. While symmetric methods, such as AES and ChaCha20, provide confidentiality, it is public key methods that often provide digital signing and key negotiation. The three most widely used methods are RSA, Elliptic Curve and ElGamal. The robustness of cryptography depends on  strong mathematical calculations that are impractical to achieve without the legit cryptographic key. A difficult calculation may take years for powerful systems to achieve.
Peter Shor, though, defined a quantum algorithm for integer factorization that runs in polynomial time \cite{Shor1995}.  With the advent of quantum computers, most of the existing public key methods will be cracked within a reasonable time limit~\cite{buchanan2017will}. There are several methods defined for quantum robust cryptography including:

\begin{itemize}

\item \textbf{Lattice-based cryptography}. This classification shows great potential and is leading to new cryptography methods, such as for fully homomorphic encryption, and code obfuscation. \item  \textbf{Code-based cryptography}. This classification was created in 1978 with the McEliece cryptosystem \cite{McEliece1978} but has barely been used in real applications. 
\item \textbf{Multivariate polynomial cryptography}. This classification involves the difficulty of solving systems of multivariate polynomials over finite fields. Unfortunately, many of the methods that have been proposed have already been broken \cite{buchanan2017will}.

\item \textbf{Hash-based signatures}. This classification involves creating random private keys and then hashing these into hash trees. The drawback is that a signer needs to keep a track of all of the messages that have been signed, and that there is a limit to the number of signatures that can be produced. 

\end{itemize}

The National Institute of Standards and Technology (NIST) runs PQC Standardization competition to improve the standards to incorporate PQC. It expects to issue the standardization papers by 2024.
This paper focuses on the coverage of the important methods that could be used within PQC, and in their evaluation around the TLS handshaking process. It includes measurements of seven key exchange methods and five signature methods. The benchmarking tests are performed within the ARMv8 64-bit architecture (aarch64), specifically a Raspberry Pi 3B+ running openSUSE, and  the open-source C-library {\em liboqs} was used to provide statistically meaningful ($n > 30$) samples.

The main contributions of this paper can be summarised, as follows:

\begin{itemize}
\item It briefly analyses the notable methods that are essential within the Post Quantum Cryptography.  
\item It provides a comparison of different KEM works to measure the key exchange. Besides, this paper provides a comparison of various signature approaches. 
\item It also introduces a set of five notions, which can be leveraged to determine the robustness of a particular algorithm. 

\end{itemize}

\section{\uppercase{Methodology}}

\subsection{Open Quantum Safe Project}

The Open Quantum Safe project (OQS, \href{https://openquantumsafe.org/}{https://openquantumsafe.org/}) is an open-source project aimed at making it easier to test new post-quantum algorithms~\cite{Stebila2017}.  Part of that provision is a C-library {\em liboqs} which is available from Github (\href{https://github.com/open-quantum-safe/liboqs}{https://github.com/open-quantum-safe/liboqs}).  The library has some post-quantum algorithms in place.

\subsubsection{Coverage}
The stated aim of the OQS project team is to provide implementations of every algorithm in the NIST round 2 competition. Tables \ref{table:key_exchange} and \ref{table:signature:type} show which ones are in the latest {\em master} branch of {\em liboqs}, at the time of writing.  Each cipher comes with parameter sets, which vary according to the NIST security level that is intended.

Within the last round (Round 3), NIST have defined the following for standardization for public-key encryption and key exchange: Classic McEliece; CRYSTALS-KYBER (Lattice); NTRU (Lattice); and SABER (Lattice). For digital signatures, the finalists are: CRYSTALS-Dilithium (Lattice); FALCON (Lattice); and Rainbow (Oil and Vinegar). These are defined as the finalists, and a winner will be chosen from these, but because CRYSTALS-Kyber, NTRU, and Saber are lattice methods, NIST only wants one winner from a lattice technique. So it has drawn up a list for an alternative of: BIKE; FrodoKEM; HQC; NTRU Prime; and SIKE for KEMs. For digital signatures,  CRYSTALS-Dilithium and FALCON are lattice methods, so an alternative list includes: GeMSS; Picnic; and SPHINCS+. NIST thus wants to guard against lattice methods being cracked in the future, and thus would like an alternative method as a backup.

\begin{table*}[!htbp]
\begin{center}
\caption{\label{table:key_exchange}NIST Round 2: Key Exchange By Type}
\begin{tabular}{|l|l|l|c|}
    \multicolumn{1}{c}{Underlying}
    & \multicolumn{1}{c}{Variant}
    & \multicolumn{1}{c}{Protocol}
    & \multicolumn{1}{c}{In {\em liboqs}?}\\
    \hline
    \multirow{9}{*}{Lattice} & \multirow{5}{*}{LWE} & Crystals-Kyber & \checked \\
    & & FrodoKEM & \checked \\
    & & LAC & $\times$ \\
    & & Saber & \checked \\
    & & Three Bears & $\times$ \\
    \cline{2-4}
    & \multirow{2}{*}{RLWE} & NewHope & \checked\\
    & & Round5 & $\times$ \\
    \cline{2-4}
    & \multirow{2}{*}{NTRU} & NTRU & \checked \\
    & & Prime NTRU & $\times$ \\
    \hline
    \multirow{7}{*}{Code-Based} & \multirow{2}{*}{Goppa} & Classic McEliece & $\times$\\
    & & NTS-KEM & $\times$\\
    \cline{2-4}
    & Moderate Density Parity-Check & BIKE & \checked \\
    \cline{2-4}
    & Hamming Quasi-Cyclic & HQC & $\times$\\
    \cline{2-4}
    & Low Density Parity-Check & LEDACrypt & $\times$\\
    \cline{2-4}
    & Low Rank Parity Check & Rollo & $\times$\\
    \cline{2-4}
    & Ideal and Gabidulin & RQC & $\times$\\
    \hline
    Isogeny & Supersingular & SIKE & \checked\\
    \hline 
\end{tabular}
\end{center}
\end{table*}


\begin{table}[htbp]
\begin{center}
\caption{\label{table:signature:type}NIST Round 2: Signature By Type}
\scalebox{.75}[.75]{
\begin{tabular}{|l|l|l|c|}
    \multicolumn{1}{c}{Underlying}
    & \multicolumn{1}{c}{Variant}
    & \multicolumn{1}{c}{Protocol}
    & \multicolumn{1}{c}{In {\em liboqs}?}\\
    \hline
    \multirow{3}{*}{Lattice} & LWE & Crystals-Dilithium & \checked \\
    \cline{2-4}
    & RLWE & qTESLA & \checked \\
    \cline{2-4}
    & NTRU & Falcon & $\times$ \\
    \hline
    \multirow{2}{*}{Hash-based} & Stateless & SPHINCS+ & \checked \\
    \cline{2-4}
    & Zero-Knowledge & Picnic & \checked \\
    \hline
    \multirow{4}{*}{Multivariate} & \multirow{2}{*}{UOV} & LUOV & $\times$\\
    & & Rainbow & \checked \\
    \cline{2-4}
    & HFEv- & GeMSS & $\times$ \\
    \cline{2-4}
    & Pure MQ & MQDSS & \checked \\
    \hline
\end{tabular}}
\end{center}
\end{table}


\subsubsection{OQS Benchmarks}
{\em liboqs} has built-in benchmarking functions for both key encapsulation and signatures.  For each cipher, these perform key operations for a set interval (default \SI{3}{\s}) and then provide the arithmetic mean and standard deviation for the time and the CPU clock cycles.

\subsection{TLS 1.3}
TLS 1.3 reduces latency as it comprises only one round trip, whereas TLS 1.2 consists of three rounds. The key encapsulation and signature schemes come together in the TLS protocol. 
To see how, we briefly outline the latest version, 1.3.    

\subsubsection{Protocol Outline}

When a client initiates a connection to a server, there are three main things that they wish to establish.  First, how are they to continue to communicate in secret; which cipher are they going to use?  Second, what is the secret that is to be used?  Third, is the server whom it claims to be?  

One feature of TLS 1.3 is that it uses an HMAC-based key derivation function (HKDF) to generate the keys from seeds.  For TLS 1.3 we need the cipher, together with the hash function, the key exchange mechanism, and the signature scheme for authentication.  RSA-based key exchanges were removed from TLS 1.3 due to security concerns.  The danger is that where encrypted traffic has been stored, a successful attempt to break the RSA scheme used, in the future, would allow that traffic to be decrypted.

\subsubsection{TLS Benchmarks}
There is a branch of the OpenSSL project into which {\em liboqs} has been integrated.  This allows us to measure the time for a TLS handshake, as follows.  Two constrained devices are linked on a simple network.  One acts as a server ({\em OpenSSL s\_server}), the other as the client ({\em OpenSSL s\_client}).  For a given set of parameters, the client establishes a connection to the server, and then terminates.  This is performed 50 times in a row, and measured with the `{\em perf}' command, yielding a good approximation of the mean time to perform the TLS handshake.

There are two dimensions under investigation.  One is the impact of using different cipher suites to sign and verify the certificates.  For this, the key exchange mechanism is held constant.  The other is the impact of varying the cipher suite used to perform the key exchange.  For this, the signing cipher is held constant. The relevant data from the output of {\em perf} are stripped out and normalised using simple bespoke scripts before being presented.

\subsection{The PQC/IoT Laboratory}

\subsubsection{Hardware}
The hardware in the laboratory is as follows.
\begin{itemize}
\item The constrained devices are the Raspberry Pi 3B+ with a \SI{1.2}{\giga\hertz} quad core ARMv8 (A53) 64-bit processor, \SI{1}{\giga\byte} RAM and \SI{32}{\giga\byte} storage.

\item Where needed, a simple standalone network is created using a home-based router (TPLink TD-W9970) as a switch.  The wired network LAN provision is 4 x 10/100~Mbps. In a real-life, IoT scenarios are likely to use low  power wireless communication technologies, such as ZigBee, BLE, LoRa, SigFox, and so on.

\item Access to the devices is over SSH from a laptop.

\item The desktop machine for compiling aarch64 code is an iMac with \SI{3}{\giga\hertz} Intel i5, with 6 cores.
\end{itemize}

\subsubsection{Software}
\begin{itemize}

\item In order to save time, a cross-compiler for ARMv8 is built and run on a more powerful desktop machine, running macOS 10.14.5.  The cross-compiler ({\em Crosstool-NG}) is built with {\em gcc} 8.3.0.  The aarch64 code is also compiled with {\em gcc} 8.3.0.

\item {\em liboqs} and OpenSSL with {\em liboqs}, are compiled for the ARMv8 (aarch64).

\item {\em liboqs} requires a 64-bit operating system.  The default operating system for a Raspberry Pi is Raspbian, which is currently 32-bit only.  We choose OpenSUSE (Leap 15.0, JeOS) since it appears to be the best maintained of the 64-bit Linux-based systems for the ARMv8 available.

\end{itemize}

\begin{figure}[htbp]
  \includegraphics[width=1.0\linewidth]{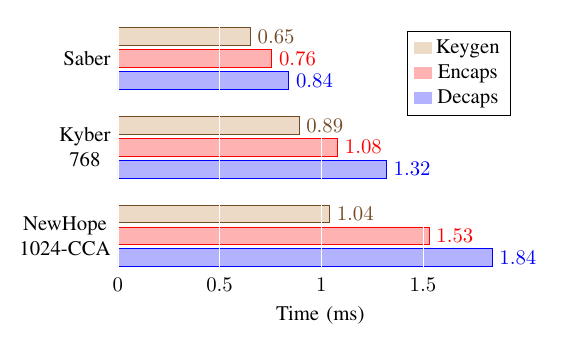} 
 \caption{KEM Meantimes: Top Performing Algorithms at Level 3}
  \label{fig:one}
\end{figure}

\begin{figure}[htbp]
  \includegraphics[width=1.0\linewidth]{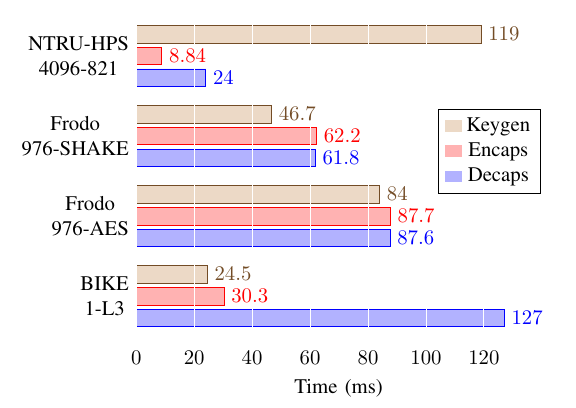} 
 \caption{KEM Meantimes: 2nd Division Algorithms at Level 3}
  \label{fig:two}
\end{figure}

\begin{figure}[htbp]
  \includegraphics[width=1.0\linewidth]{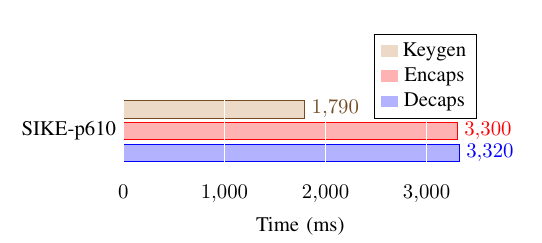} 
 \caption{KEM Meantimes: 3rd Division Algorithm at Level 3}
  \label{fig:three}
\end{figure}



\section{Results}\label{sec:results}

In this paper, we present our results and compare them, where possible, with other benchmarks from the literature.  First, we look at the key exchange mechanisms.  That is followed by the results for the signature schemes.   We then look at the TLS handshake which combines both exchange mechanisms and signatures. Finally, we offer some considerations which aim to help with choosing a reliable cryptographic algorithm.

\subsection{Key Exchange Mechanisms}

To gain an appreciation for the different systems, we chart the public and private key sizes in Table~\ref{table:key:exchange:submitted:keys}.  These are gleaned from the second round submissions for: Crystals-Kyber~\cite{Avanzi2019a}, FrodoKEM~\cite{Alkim2019}, SABER~\cite{DAnvers2019},  NewHope~\cite{Poppelmann2019}, NTRU~\cite{Chen2019}, BIKE~\cite{Aragon2019} and SIKE~\cite{Jao2019}.


In order that the numbers are not overwhelming, we restrict the view in several ways.  First, we choose the parameter set which has been provided for achieving level three security, which is equivalent to breaking AES-256, or alternatively 128 bits of quantum security.  Second, where IND-CPA and IND-CCA versions are specified, we restrict the view to IND-CCA.  Third, we pick the first BIKE model as representative of that family of three implementations.  Lastly, we only list those which currently have a representation in {\em liboqs}.  These restrictions overlook many subtleties, but nevertheless yield insight.

%
%
\begin{table*}
\begin{center}
\caption{\label{table:key:exchange:submitted:keys}Key Exchange Submitted Public and Private Key Length (Bytes)}
\begin{tabular}{|l|l|l|r|r|}
      \multicolumn{1}{c}{Underlying}
    & \multicolumn{1}{c}{Variant: Protocol}
    & \multicolumn{1}{c}{Parameter Set}
    & \multicolumn{1}{c}{Private}
    & \multicolumn{1}{c}{Public}\\
    \hline
    \multirow{5}{*}{Lattice} & LWE: Saber & SABER-KEM & 2,304 & 992 \\
    & LWE: Crystals-Kyber & Kyber-768 & 2,400 & 1,184\\
    & LWE: FrodoKEM & FrodoKEM-976 & 31,296 & 15,632 \\
    \cline{2-5}
    & RLWE: NewHope & NewHope1024  & 3,680 & 1,824\\
    \cline{2-5}
    & NTRU: NTRU & ntruhps4096821 & 1,592 & 1,230 \\
    \hline
    Code-Based & MDPC: BIKE & BIKE-1-CCA & 6,592 & 6,205 \\
    \hline
    Isogeny & Supersingular: SIKE & SIKEp610 & 524 & 462 \\
    \hline 
\end{tabular}
\end{center}


\end{table*}

With respect to speed, the candidates fall into three divisions, with lattice-based solutions being the front-runners.  In the first division are Saber, Kyber and NewHope.  Saber has both a reasonable key size and is performing its operations sub-millisecond.  The second division is made up of Frodo, NTRU and code-based BIKE.  Finally, SIKE is in a division of its own; whilst having a reasonable key size, it is taking over 3 full seconds for encapsulating or decapsulating.  These divisions are compared in Figures~\ref{fig:one}, \ref{fig:two} and \ref{fig:three}.

\subsubsection{Comparison with Other Work}

Table~\ref{table:comparison:others:kem} shows the results from three other papers to measure the key exchange for post-quantum algorithms.  The listed speed is the speed of an encapsulation followed by a decapsulation.  The table presents the best attempt to provide a like-for-like comparison between the various works. The main point is that the benchmarking is conducted under differing assumptions.  
 A brief outline of the works illustrates the differences.

\begin{table*}
\begin{center}
\caption{\label{table:comparison:others:kem}Comparison with Other KEM Work}
\begin{tabular}{|l|l|r|c|r|}
      \multicolumn{1}{c}{Cipher}
    & \multicolumn{1}{c}{Bench}
    & \multicolumn{1}{c}{Security}
    & \multicolumn{1}{c}{Constrained?}
    & \multicolumn{1}{c}{Speed }\\
    & \multirow{2}{*}{} & Level & & (ms) \\
    \hline
    Saber  &      \cite{Karmakar2018}  & ---  & \checked  &  18.84  \\
    Saber  &      \cite{Kannwischer2019}  & 185  & \checked  &  98.60   \\
    \hline
    Saber  &      This Work      & 185 & \checked   &  1.60  \\
    \hline
    \hline
    Kyber  &      \cite{An2018}  & ---  & $\times$  &   0.38  \\
    Kyber  &      \cite{Kannwischer2019}  & 164  & \checked  &   93.39  \\
    \hline
    Kyber  &      This Work      & 164    & \checked   &   2.40 \\
    \hline
    \hline
    NewHope &     \cite{An2018}      &  233  & $\times$  &   0.23 \\
    NewHope &     \cite{Malina2018}  &  206  &  \checked &      2.36 \\
    NewHope &     \cite{Kannwischer2019}  &  233  &  \checked  &  159.61  \\
    \hline
    NewHope  &     This Work         &   233 &   \checked &   3.38   \\
    \hline
    \hline
    NTRU    &      \cite{An2018}     &  ---  & $\times$  &    2.06 \\
    NTRU    &      \cite{Malina2018} &  128  & \checked  &   12.88 \\
    NTRU    &     \cite{Kannwischer2019}  &  128  &  \checked &  94.69  \\
    \hline
    NTRU     &     This Work         &  128  & \checked  & 32.82 \\
    \hline
    \hline
    Frodo   &      \cite{An2018}     &  ---  & $\times$  &      9.38 \\
    Frodo   &      \cite{Malina2018} &  130  & \checked  &   702.12 \\
    \hline
    Frodo    &     This Work         &  150 & \checked   &   175.30    \\
    \hline
    \hline
    SIKE    & \cite{Seo2019}     & 192 &  \checked   &    303.00 \\
    SIKE   &     \cite{Kannwischer2019}  & 192  &  \checked &      279,866.01 \\
    \hline
    SIKE   &  This Work          & 192 &  \checked   &   6,620.41  \\
    \hline 
\end{tabular}
\end{center}

\end{table*}


Seo et al. \cite{Seo2019} implement a highly optimised version of SIKE, written in assembly for the aarch64 instruction set, and tested on an ARMv8 (A53) running 64-bit code at \SI{1.536}{\giga\hertz}.   For SIKEp610 (NIST level 3), the reported clock cycles (crudely) equate to \SI{303}{\milli\second}.  An et al. \cite{An2018} measure the performance of the protocols then available in {\em liboqs}: Frodo, BCNS, NewHope, MSrln, Kyber, NTRU, McBits, IQC and SIDH.  The tests are on an Intel i7-5500 at \SI{2.4}{\giga\hertz}, two cores, with \SI{16}{\giga\byte} RAM.  For Kyber and Frodo, they use non-standard parameter sets for which the post-quantum security level is unclear.  

Malina et al. \cite{Malina2018} test six key exchange mechanisms; NewHope, NTRU, BCNS, Frodo, McBits and SIDH.  
To do this they tweaked {\em liboqs} to run in 32-bits, and then ran tests on (a) an Android 6 device with  Qualcomm Snapdragon 801 (4 cores) at \SI{2.5}{\giga\hertz} with \SI{2}{\giga\byte} RAM, and (b) a Raspberry Pi 2, running Raspbian Strech Lite, on an ARMv7 at \SI{1.2}{\giga\hertz}, with \SI{1}{\giga\byte} RAM. 
Karmakar et al. \cite{Karmakar2018} implement an optimised Saber on an ARM Cortex-M4, with an assumed clock of \SI{168}{\mega\hertz}. Kannwischer et al. \cite{Kannwischer2019} \label{kem:kannwischer:explained} from July 2019 present a benchmarking framework {\em pqm4}.  
They consistently benched 10 key encapsulation and five signature schemes on a 32-bit ARM Cortex-M4 running at \SI{24}{\mega\hertz}, with \SI{196}{\kilo\byte} of RAM.  Frodo-976 takes too much resource to run in this environment.

There are several other related works which are not suitable for inclusion in Table~\ref{table:comparison:others:kem}, since they are even more disparate than those included.

\begin{itemize}

\item \cite{Czypek2012} looks at MQ-based systems on an AtxMega128a1 microprocessor, running at \SI{32}{\mega\hertz}, with \SI{8}{\kilo\byte} RAM and \SI{128}{\kilo\byte} storage.

\item \cite{Johansson2016} compares the speed of an RLWE cipher against classical ciphers, when used for the key exchange during TLS handshakes.  Benchmarks are from an ARM Cortex-A7 running at \SI{700}{\mega\hertz} with \SI{1}{\giga\byte} of RAM and  storage of over \SI{1}{\giga\byte}.  They classify the classical security level at \SI{128}{bit}.  

\item \cite{Suomalainen2018} uses the ABSOLUT~\cite{Vatjus-Anttila2013} tool to test Frodo and NewHope on simulated (a) Raspberry Pi 2,  with Broadcom BCM2836 at \SI{900}{\mega\hertz} quad-core 32-bit ARM Cortex-A7 processor with \SI{1}{\giga\byte} RAM, and (b) Raspberry Pi 3, with Broadcom BCM2837 at \SI{1.2}{\giga\hertz} quad-core 64-bit ARM Cortex-A53, also with \SI{1}{\giga\byte} RAM.  

\item \cite{Crockett2019} provides extensive observations on the integration of post-quantum algorithms into TLS and SSH.

\end{itemize}

\subsection{Signatures}

For the different signature schemes, we chart the public and private key sizes together with the signature size, in Table~\ref{table:signature:submitted:keys}.  These are taken from the second round submissions for: Dilithium~\cite{Ducas2019}, qTESLA~\cite{Bindel2019},  MQDSS~\cite{Chen2019a}, Rainbow~\cite{Ding2019} and SPHINCS+~\cite{Aumasson2019}.  As for key exchanges, we restrict the view to those parameter sets which aim for security at NIST level 3, and only list those which currently have a representation in {\em liboqs}.

\begin{table*}[htbp]
\begin{center}
\caption{\label{table:signature:submitted:keys}Signature Schemes: Key and Signature Lengths (Bytes)}
\begin{tabular}{|l|l|r|r|r|}
      \multicolumn{1}{c}{Underlying}
    & \multicolumn{1}{c}{Parameter Set}
    & \multicolumn{1}{c}{Private}
    & \multicolumn{1}{c}{Public}
    & \multicolumn{1}{c}{Signature} \\
    \hline
    Lattice (LWE)  & Dilithium IV  & 96   &   1,760 &  3,366 \\
    \hline
    Lattice (RLWE)   & qTESLA-III   &  2,368  &   3,104 &   2,848 \\
    \hline
    \multirow{3}{*}{MQ} & MQDSS-31-64 &   24   &   64  & 43,728 \\
                        & Rainbow-IIIc &  \SI{511.4}{\kilo\byte}  & \SI{710.6}{\kilo\byte}  & 156 \\
                        & Rainbow-IIIc-cyclic &   \SI{511.4}{\kilo\byte} & \SI{206.7}{\kilo\byte}  & 156 \\
    \hline
    \multirow{2}{*}{Hash-based}    & SPHINCS+-192s & 96  &  48  &   17,064 \\
                                   & SPHINCS+-192f &  96  &  48 &   35,664 \\
\hline
\end{tabular}
\end{center}

\end{table*}

\begin{figure}[htbp]
  \includegraphics[width=1.0\linewidth]{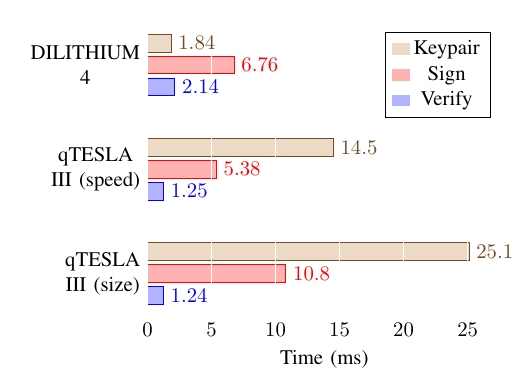} 
  \caption{Signature Meantimes: Lattice-based at Level 3}
  \label{fig:four}
\end{figure}
\begin{figure*}
  \includegraphics[width=0.8\textwidth]{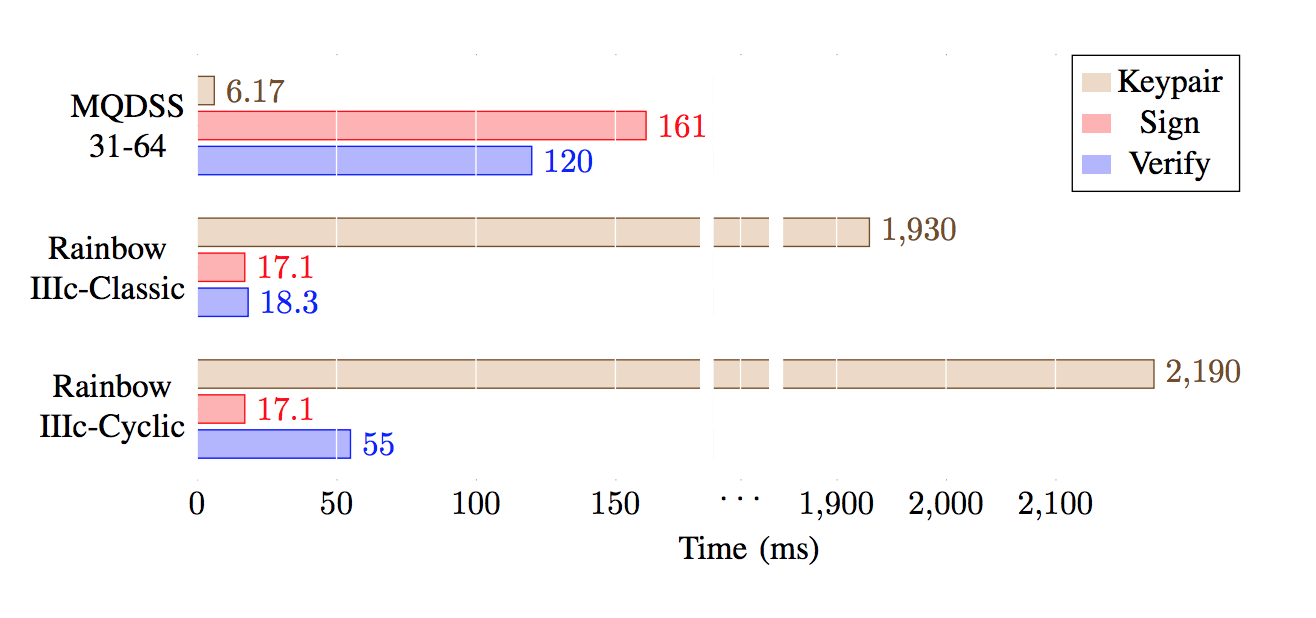} 
  \caption{Signature Meantines: MQ-based at Level 3}
  \label{fig:five}
\end{figure*}
\begin{figure*}
  \includegraphics[width=0.8\textwidth]{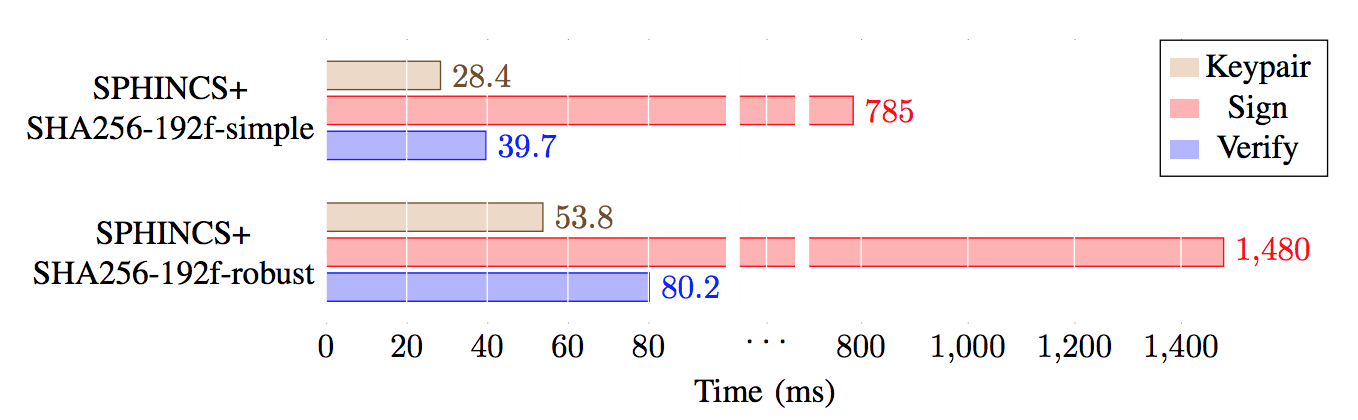} 
  \caption{Signature Meantimes: SPHINCS+ `f' at Level 3}
  \label{fig:six}
  \vspace{-0.5cm}

\end{figure*}
\begin{figure*}
  \includegraphics[width=0.8\textwidth]{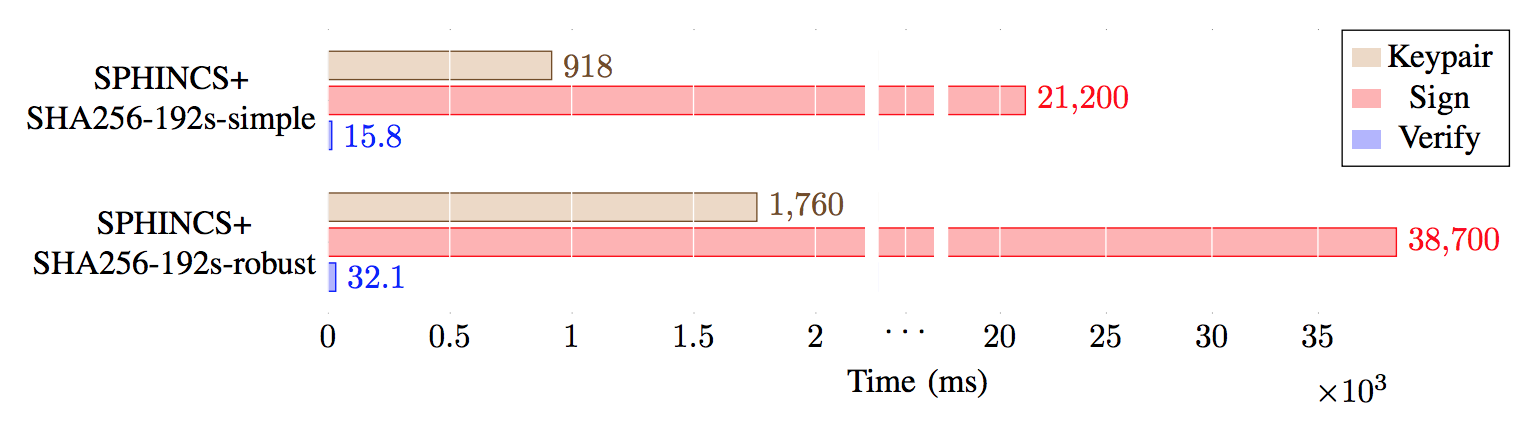} 
  \caption{Signature Meantimes: SPHINCS+ SHA256 `s' at Level 3}
  \label{fig:seven}

\end{figure*}

Figures~\ref{fig:four}, \ref{fig:five}, \ref{fig:six} and \ref{fig:seven} show how they line up against each other.  There are some items which should be taken into account with these comparisons.

\paragraph{qTESLA} There is a version of qTESLA which is probably secure, but it is not yet implemented in {\em liboqs} and has not been tested here.

\paragraph{SPHINCS+} SPHINCS+ presents three sets of parameters of various sorts.  First, since it is hash-based it needs a hash function.  There are three options: Haraka, SHA256 and SHAKE256.  Second, they produce a set of parameters optimized for small signatures (`s' for small) and fast processing (`f' for fast).  

For a given NIST level, there are therefore 12 different sets of parameters.  We take SHA256 to be representative.

\paragraph{Rainbow} The `compressed' version can take the private signature down to 64 bytes (512 bits).

\subsubsection{Comparison with Other Work}

Table~\ref{table:comparison:others:sig} shows a comparison between the benchmarks from \cite{Kannwischer2019} and the results presented here.  The speed is the total speed of a sign/verify pair.

\begin{table*}[htbp]
\begin{center}
\caption{\label{table:comparison:others:sig}Comparison with Other Signature Work}
\begin{tabular}{|ll|l|r|c|r|}
    \multicolumn{2}{c}{Cipher} &
    \multicolumn{1}{c}{Bench}  &
    \multicolumn{1}{c}{Speed (ms)}\\
    \hline
    \multirow{2}{*}{Dilithium IV} & & \cite{Kannwischer2019} & 485.10\\
               & & This Work & 0.01\\
    \hline
    \multirow{2}{*}{qTESLA-III} &  & \cite{Kannwischer2019} & 354.06\\
               &  & This Work & 2.70 \\
    \hline           
    \multirow{8}{*}{SPHINCS+-SHA256-192} & f-sim & \cite{Kannwischer2019} &  30,116.29  \\
    & f-sim & This Work &  824,908.31  \\
    \cline{3-4}
    & f-rob &  \cite{Kannwischer2019} &  57,169.86 \\   
    & f-rob &  This Work &  1,558,082.26 \\
    \cline{3-4}   
    & s-sim &  \cite{Kannwischer2019} &  793,925.00  \\
    & s-sim &  This Work &   21,187,279.39 \\
    \cline{3-4}
    & s-rob &  \cite{Kannwischer2019} &  1,433,837.19 \\   
    & s-rob &  This Work &  38,682,673.41 \\   
\hline
\end{tabular}
\end{center}

\end{table*}

\subsection{TLS 1.3 Handshake}
In our tests, varying the signature algorithm made no difference to the size of the handshake (\SI{2041}{\byte} read, \SI{391}{\byte} written), nor to the (wall-clock) speed of the handshake (mean time of \SI{0.05}{\second}).  This may be because the two PQC signature algorithms available, Dilithium and qTESLA, are both lattice-based.  One area for further work is to examine this area in more detail, when further algorithms are available.  The importance of the security of the signature algorithm should not be underestimated; it ensures the authentication of the counterparty.

Table~\ref{table:tls:handshake:by:kem} shows how the size and speed of the TLS 1.3 handshake varies with the key exchange.  This shows that, as we expect from the results above, that Saber, Crystal and NewHope are light and fast; BIKE, Frodo and NTRU are comparable.  While the handshake for SIKE is the smallest of all, it takes orders of magnitude longer.

%
\begin{table*}
\begin{center}
\caption{\label{table:tls:handshake:by:kem}TLS Handshake Profiles by KEM}
\scalebox{.90}[.90]{
\begin{tabular}{|l|l|r|r|r|r|}
      \multicolumn{2}{c}{}
    & \multicolumn{3}{c}{Handshake (Bytes)}
    & \multicolumn{1}{c}{Speed} \\
      \multicolumn{1}{c}{Protocol}
    & \multicolumn{1}{c}{Parameter Set} 
    & \multicolumn{1}{c}{Read} 
    & \multicolumn{1}{c}{Write} 
    & \multicolumn{1}{c}{Total} 
    & \multicolumn{1}{c}{(\SI{}{\milli\second})}\\
    \hline
    Crystals-Kyber & Kyber-768 & 3,897 & 1,543 & 5,440 & 89.82 \\
    \hline
    FrodoKEM & FrodoKEM-976 & 18,553 & 15,991 & 34,544 & 295.81\\
    \hline
    Saber & SABER-KEM          & 3,897 & 1,351 & 5,248 & 87.69 \\
    \hline
    NewHope & NewHope1024     & 5,017 & 2,183 & 7,200 & 88.12 \\
    \hline
    NTRU & ntruhps4096821     & 4,039 & 1,589 & 5,628 & 241.34 \\
    \hline
    BIKE & BIKE-1-CCA         & 7,773 & 5,323 & 13,096 & 199.66 \\
    \hline
    SIKE & SIKEp610  & 3,295 & 821  & 4,116 & 8,580.58 \\
    \hline 
\end{tabular}}
\end{center}

\end{table*}

\subsubsection{Comparison with Other Work}

Table~\ref{table:comparison:others:tls} shows two other attempts from the literature to measure 
post-quantum key exchange in TLS handshakes.  The same caveats apply as for the comparisons in Table~\ref{table:comparison:others:kem}.

\begin{table*}[htbp]
\begin{center}
\caption{\label{table:comparison:others:tls}Comparison with Other TLS Work}
\begin{tabular}{|l|l|r|c|r|r|}
    \multicolumn{1}{l}{Cipher} &
    \multicolumn{1}{l}{Bench}  &
    \multicolumn{1}{r}{Security Level} &
    \multicolumn{1}{c}{Constrained?}  &
    \multicolumn{1}{r}{Handshake (Bytes)}  &
    \multicolumn{1}{r}{Speed (ms)}\\
    \hline
    NewHope &     \cite{Bos2016}     &  206  & $\times$  &  5,514   &    13.10 \\
    RLWE    &     \cite{Bos2015}     &  82   & $\times$  &  10,479  &    54.00 \\
    \hline
    NewHope  &     This Work         &   233 &   \checked &   7,200  &     88.12 \\
    \hline
    \hline
    NTRU    &      \cite{Bos2016}    &  128  & $\times$  & 3,691   &  19.90 \\
    \hline
    NTRU     &     This Work         &  128  & \checked  & 5,628   & 241.34 \\
    \hline
    \hline
    Frodo   &      \cite{Bos2016}    &  130  & $\times$  &  24,228     &    20.70 \\
    \hline
    Frodo    &     This Work         &  150 & \checked   &  34,554     &     295.81 \\
    \hline
\end{tabular}
\end{center}


\end{table*}

\paragraph{\cite{Bos2015}} implements a RLWE ciphersuite in OpenSSL (v1.0.1f) and benchmarks interactions with an Apache webserver.   The parameters they choose aim for 128-bit security, which matches the rationale chosen here.  They provide results for retrieving a 1 byte payload, which is comparable with completing a TLS handshake.  One difference is that they code with TLS 1.2, which is less efficient than TLS 1.3.  The largest difference is that their devices are not constrained: the `client' machine is based on an Intel i5 with 4 cores running at \SI{2.7}{\giga\hertz}; the `server' machine is an Intel Duo with 2 cores running at \SI{2.33}{\giga\hertz}.

\paragraph{\cite{Bos2016}} benchmarks connections to a TLS-protected webserver for Frodo, BCNS, NewHope and NTRU.  These were run on a Google Cloud VM (n1-standard-4), with 4 virtual CPUs at \SI{2.6}{\giga\hertz}, with \SI{15}{\giga\byte} of RAM.  The clients were similarly specified (n1-standard-32).  The parameters were aimed at 128 bits of security.  Some testing was also done on a BeagleBone Black, with an AM335x ARM Cortex-A8 running at \SI{1}{\giga\hertz}. The BeagleBone was not running a TLS client, and there is no available comparison on that front with this work.

\subsection{Choosing an Algorithm}

The purpose of the encryption is an important consideration. Suppose the target is to verify a system update for an IoT device.  Then the signing will be done by the system issuing the update; assume this is not itself IoT.  In that case the SPHINCS+ suite might be a viable choice.  The IoT device only needs to verify the signature, which we have shown it can do sub-second (at level 3).  That may well be fast enough for relatively infrequent system updates.  On the other hand, SIKE (SIKE-p610) would be inappropriate for involvement in the key exchanges of TLS since the encapsulation followed by decapsulation would add over six seconds to the handshake.  So whether an algorithm is `fast enough' will depend on the use to which it is being put.  Similar considerations will apply to the public/private key size. 

Of the NIST key exchange finalists, there are three main security levels (1, 3 and 5), and the key sizes for each method is defined in Table \ref{kem2} \cite{asecuritysite}. It can be seen that McEliece requires relatively large private keys, and fairly large public keys. The lattice methods (Kyber, SABER and NTRU) offer a good balance for key sizes and ciphertext compared with McEliece. With digital signatures, as shown in Table \ref{kem3}, shows that SPHINCS+ supports the smallest key sizes, and the largest digital signature size. The lattice methods, again, offer a good compromise, and where the Oil and Vinegar method (Rainbow) has fairly large key sizes, but the smallest digital signature size.

\begin{table*}
\begin{center}
\caption{\label{kem2} Key sizes and ciphertext for Key exchange methods }
\begin{tabular}{|l|l|r|r|r|r|}    \hline
\textbf{Method} & \textbf{Public key (B)} & \textbf{Private key (B)} & \textbf{Ciphertext (B)}\\
    \hline
Kyber512     &         800      &        1,632       &           768\\    \hline
Kyber738     &       1,184      &        2,400       &         1,088\\    \hline
Kyber1024     &      1,568       &       3,168       &         1,568\\    \hline
LightSABER    &        672        &      1,568       &           736\\    \hline
SABER         &        992       &       2,304        &        1,088\\    \hline
FireSABER     &      1,312       &       3,040        &        1,472\\    \hline
McEliece348864 &   261,120       &       6,452        &          128\\    \hline
McEliece460896 &   524,160       &      13,568         &         188\\    \hline
McEliece6688128& 1,044,992       &      13,892        &          240\\    \hline
McEliece6960119& 1,047,319       &      13,948        &          226\\    \hline
McEliece8192128& 1,357,824       &      14,120        &          240\\    \hline
NTRUhps2048509 &       699       &         935        &          699\\    \hline
NTRUhps2048677  &      930       &       1,234        &          930\\    \hline
NTRUhps4096821   &   1,230       &       1,590        &        1,230   \\ 
    \hline 
\end{tabular}
\end{center}

\end{table*}

\begin{table*}
\begin{center}
\caption{\label{kem3} Key sizes and signature size for Digital Signature methods}
\begin{tabular}{|l|l|r|r|r|r|}    \hline
\textbf{Method} & \textbf{Public key (B)} & \textbf{Private key (B)} & \textbf{Signature  (B)}\\
    \hline
Crystals Dilithium 2  &      1,312      &        2,528         &     2,420   \\   \hline
Crystals Dilithium 3        &          1,952     &         4,000  &            3,293  \\   \hline  
Crystals Dilithium 5        &          2,592       &       4,864       &       4,595    \\   \hline 
Falcon 512 (Lattice)         &           897       &       1,281      &          690   \\ \hline
Falcon 1024          &                 1,793       &       2,305     &         1,330    \\ \hline
Rainbow Level Ia &  161,600        &    103,648   &              66    \\ \hline
Rainbow Level IIIa      &            861,400     &       611,300   &              164      \\ \hline
Rainbow Level Vc        &          1,885,400     &     1,375,700        &        204       \\ \hline
SPHINCS+ SHA256-128      &          32      &           64    &          17,088   \\ \hline
SPHINCS+ SHA256-192    &            48        &         96   &          35,664     \\ \hline
SPHINCS+ SHA256-256    &            64        &        128    &          49,856     \\ 

    \hline 
\end{tabular}
\end{center}

\end{table*}

Table~\ref{table:consid:certainty:security} summarises the important elements involved for the key considerations for post-quantum cryptography. An important element is how \emph{venerable} the underlying mathematical problem, and where the  more cryptanalysis it has survived,  the better. Another important element is that the method must be \emph{$\mathit{\mathbf{NP}}$-hard}, and have a related proof with a \emph{problem-reduction}. With QROM (quantum random oracle model)-secure, we have a proof of security in the QROM, along with ROM (random oracle model))-secure. The existence of such proof does not entail security against a classical adversary, but it is taken to be a good heuristic. (\cite[6]{Ducas2019} conjectures that whilst the theoretical details differ, security in the ROM and security in the QROM will converge to indicate the same level of security in practice.)

\begin{table*}[htbp]
\begin{center}
\caption{\label{table:consid:certainty:security}Considerations for Certainty about Security}
\begin{tabular}{l|l|rcccc}
    \multicolumn{1}{c}{Scheme} &
    \multicolumn{1}{c}{Underlying} &
    Venerability (yrs) &
    $\mathit{\mathbf{NP}}$-hard &
    Problem-Reduction &
    ROM-secure &
    QROM-secure \\
    \hline
    Saber     & Lattice: LWE  & 14$^{a}$  & \checked$^{i}$ &  $\times^{i}$  & \checked$^{q}$ & \checked$^{q}$ \\ 
    Kyber     & Lattice: LWE  & 14$^{a}$  & \checked$^{i}$ &  $\times^{i}$  & \checked$^{r}$ & \checked$^{r}$ \\
    Frodo     & Lattice: LWE  & 14$^{a}$  & \checked$^{i}$ &  $\times^{i}$  & \checked$^{s}$ & \checked$^{s}$ \\
    NewHope   & Lattice: RLWE & 9$^{b}$   & \checked$^{j}$ &  $\times^{j}$  & \checked$^{t}$ & \checked$^{t}$\\
    NTRU      & Lattice: NTRU & 23$^{c}$  & \checked$^{k}$ &  $\times^{k}$  & \checked$^{u}$ & \checked$^{u}$\\ 
    BIKE      & Code: MDPC    & 41$^{d}$  & \checked$^{l}$ &  \checked$^{n}$ & \checked$^{n}$ & $\times^{n}$ \\
    SIKE      & Supersingular &  5$^{e}$  & $\times$ & \checked$^{v}$ &  \checked$^{v}$ & $\times^{v}$ \\ 
    \hline
    \hline
    Dilithium & Lattice: LWE    & 14$^{a}$ & \checked$^{i}$ & $\times^{i}$ & \checked$^{w}$ & \checked$^{w}$ \\
    qTESLA    & Lattice: RLWE   & 9$^{b}$  & \checked$^{j}$ & $\times^{j}$ & \checked$^{x}$ & \checked$^{x}$ \\
    MQDSS     & MQ              & 31$^{f}$ & \checked$^{m}$ & \checked$^{o}$ &  \checked$^{o}$  & $\times^{o}$ \\ 
    Rainbow   & MQ: UOV         & 20$^{g}$ & \checked$^{m}$ & $\times^{p}$ & $\times^{y}$  & $\times^{y}$\\
    SPHINCS+  & Hash: Haraka    & 4$^{h}$  & --- &  --- & --- & ---\\
    SPHINCS+  & Hash: SHA256    & 18$^{h}$ & --- &  --- & --- & ---\\
    SPHINCS+  & Hash: SHAKE256  & 7$^{h}$  & --- &  --- & --- & ---\\
    \hline
\end{tabular}
\end{center}

\scriptsize

Notes:
\begin{multicols}{2}
\begin{itemize}
\item[a] 2005 \cite{Regev2009} is a revision of the original.
\item[b] 2010 \cite{Lyubashevsky2010}.
\item[c] 1996 \cite{Ajtai1996}.
\item[d] 1978 \cite{McEliece1978} \cite{Misoczki2013}.
\item[e] 2014 \cite{DeFeo2014}.
\item[f] 1988 \cite{Matsumoto1988}.
\item[g] 1999 \cite{Kipnis1999}.
\item[h] SPHINCS+ is no more secure than the hash it uses~\cite[41]{Butin2017}.  Haraka dates from 2015~\cite{Kolbl2017}; SHA256 from 2001~\cite{NIST2002}; SHAKE256 from 2012 (see~\cite{NIST2015}). The notions of NP-hardness, or security in ROM do not apply directly to hashes.
\item[i] SVP and variants have been proved NP-hard \cite{Micciancio2002}.  However, the systems rely on approximations which have not \cite{Peikert2016}, hence no Problem Reduction.
\item[j] RLWE is NP-hard \cite{Peikert2016}; however, as for (i), the systems rely on approximation problems which have not.
\item[k] The NTRU problem is not reducible to a lattice problem \cite[14]{Peikert2016}.  However, RLWE is at least as hard as NTRU and if RLWE is hard, then NTRU can be made secure \cite[14]{Peikert2016}.
\item[l] \cite{Berlekamp1978} 
\item[m] \cite{Garey1979} 
\item[n] \cite[60-1]{Aragon2019} 
\item[o] \cite{Chen2016b} 
\item[p] \cite[43]{Ding2019} 
\item[q] \cite[12-3]{DAnvers2019}
\item[r] \cite[19-20]{Avanzi2019a}
\item[s] \cite[31-5]{Alkim2019}
\item[t] \cite[28-29]{Poppelmann2019}
\item[u] \cite[30]{Chen2019}
\item[v] \cite[41-2]{Jao2019}
\item[w] \cite[5-6]{Ducas2019}
\item[x] \cite[47-8]{Bindel2019}
\item[y] \cite[43]{Ding2019}
\end{itemize}
\end{multicols}


\end{table*}

\section{\uppercase{Conclusions}}\label{sec:conclusions}

In terms of KEM performance, SABER is the clear winner, executing its basic operations in under a millisecond; the TLS handshake in under \SI{90}{\milli\second}.  At the other end, SIKE's performance makes it an unlikely choice; the empty TLS handshake taking over \SI{8}{\second}.  It is true that there is at least one finely-tuned version written in assembly which performs better.


\bibliographystyle{apalike}
{\small
\bibliography{example}}

\end{document}